\documentclass[twocolumn,showpacs,nofootinbib,prd,10pt,amssymb]{revtex4-2}
\usepackage{latexsym}
\usepackage{graphicx,epsf, epsfig, amssymb, amsmath}
\usepackage{relsize}
\usepackage{dsfont}
\usepackage{todonotes}
\usepackage{textcomp}
\usepackage{float}

\DeclareMathAlphabet\mathbfcal{OMS}{cmsy}{b}{n}

\newcommand{\aap}{Astron.\ Astrophys.}
\newcommand{\mnras}{Mon.\ Not.\ R.\ Astron.\ Soc.}

\begin{document}

\title{Nonequilibrium thermodynamics of accreted neutron-star crust}

\author{
	Mikhail~E.~Gusakov,
	Elena~M.~Kantor, 
	and Andrey~I.~Chugunov
} 
\affiliation{Ioffe Institute, Polytekhnicheskaya 26, 194021 Saint-Petersburg, Russia}
\begin{abstract} 
We show that, 
in order to determine the equation of state of the inner crust of an accreting neutron star,
one should minimize not the Gibbs free energy, 
as it is generally assumed in the literature,
but a different thermodynamic potential $\Psi$,
which tends to the minimum at fixed pressure and neutron chemical potential.
Once this potential is specified,
one can calculate the heat-release distribution 
in the stellar crust due to nonequilibrium nuclear reactions
induced by accretion of matter onto the neutron-star surface.
The results are important for adequate modeling 
of the accreted crust 
and 
interpretation of the observations of accreting 
neutron stars in low-mass X-ray binaries.
\end{abstract}
\date{\today}

\pacs{	}


\maketitle

\section{Introduction}
Observations of accreting neutron stars (NSs) may shed light 
on the properties of superdense matter in their interiors 
(e.g., 
\cite{hpy07,ch08,hs17,Fortin_ea18,Brown_ea18, mdkse18,pcc19,Parikh_ea19,Wijngaarden_ea20,pc21,Fortin_ea21,sgc21}).
To interpret the observations, one needs to know the equation of state of accreted crust and 
the heat release due to nonequilibrium nuclear 
reactions 
induced by accretion.

Previously, the NS crust was studied within the so-called {\it traditional approach},
which assumes that 
crustal matter sinks deeper and deeper inside the star {\it as a whole} 
under the weight of newly accreted material \cite{sato79,HZ90,HZ90b,HZ03,Gupta_ea07, HZ08,Steiner12,Chamel_etal15_Drip,lau_ea18,Fantina_ea18,SC19_MNRAS}.
This
approach  
can be applied to
the outer crust,
consisting of atomic nuclei and electrons.
However, in the inner crust, where unbound neutrons are additionally present,
the approach is not justified, 
since it leads to violation of hydrostatic (and diffusion) 
equilibrium 
condition
for neutrons \cite{GC19_DiffEq} (hereafter, the nHD condition):
\begin{align}
\mu_{n}^\infty\equiv \mu_{n} {\rm e}^{\nu(r)/2}={\rm const},
\label{nHDcond}
\end{align}
where $\mu_{n} $ is the neutron chemical potential;
${\rm e}^{\nu(r)/2}$ is the redshift factor; 
$r$ is the radial coordinate.
Here and below, the superscript $^\infty$ indicates
that the corresponding quantity (e.g., $\mu_{n}$) is redshifted.
In major part of the crust region,
where neutrons are superfluid,
the nHD condition (\ref{nHDcond}) 
is required for hydrostatic equilibrium;
in the narrow layer $\lesssim 5$~m near the outer-inner crust interface (oi interface),
where neutrons can be nonsuperfluid,
(\ref{nHDcond}) is necessary 
for establishing
the neutron diffusion equilibrium
(one can check that neutron diffusion is very efficient 
in this layer of the star \cite{GC19_DiffEq}).
One way or the other, unbound neutrons can move independently of nuclei 
in order to decrease the system energy, and the nHD condition 
represents this fact.

How does the condition (\ref{nHDcond}) may affect 
calculations of the inner crust structure for accreting NSs?
Generally,
to derive the equation of state of the NS crust in the traditional approach,  
one 
minimizes the Gibbs free energy (e.g., \cite{HZ90,HZ90b,HZ03,HZ08,Steiner12,Chamel_etal15_Drip,Fantina_ea18,SC19_MNRAS}).
But what thermodynamic potential should be minimized 
in order to respect the condition (\ref{nHDcond})?
The Letter aims to answer this question.
As a by-product of this work, we shall also demonstrate 
how to calculate the distribution of the heat release in the accreting crust
for any given crustal equation of state.

\section{Thermodynamically consistent accreted crust}
\label{nHd}

Combining the condition (\ref{nHDcond}) with the hydrostatic equilibrium equation,
\begin{align}
\frac{dP}{dr}=-\frac{1}{2}(\varepsilon+P) \frac{d \nu}{dr},
\label{hydro}
\end{align}
yields \cite{GC19_DiffEq}:
\begin{align}
\frac{d\mu_{n}}{\mu_{n}}=\frac{dP}{\varepsilon+P}.
\label{mun}
\end{align}
Here $P$ and $\varepsilon$ are the pressure and energy density, respectively. 
Integrating (\ref{mun}) from the oi interface downward to the stellar core, one gets
\begin{align}
\mu_{n}=m_{n} \, {\rm exp}\{\int_{P_{\rm oi}}^P d\widetilde{P}/[\varepsilon(\widetilde{P})+\widetilde{P}]\},
\label{mun2}
\end{align}
where $P_{\rm oi}$ is the pressure at the oi interface 
and we took into account that $\mu_{n}=m_{n}$ there
($m_n$ is the bare neutron mass; here and below the speed of light $c$ is set to unity).
Equation (\ref{mun2}) is very important; 
it says that $\mu_n$ at any point in the inner crust with the pressure $P$
is uniquely determined once $P_{\rm oi}$ and the function $\varepsilon(\widetilde{P})$  are specified 
for $P_{\rm oi}\leq \widetilde{P}< P$.

It is easy to find the pressure $P_{\rm cc}$ at the crust-core boundary
by matching
$\mu_{n}$ at 
the boundary,
$\mu_{n}(P_{\rm cc})=\mu_{n, {\rm core}}(P_{\rm cc})$,
where $\mu_{n, {\rm core}}(P)$ is the neutron chemical potential in the core, 
a known function of pressure.  

In the course of accretion, the NS crust eventually enters 
the regime,
in which
further accretion almost does not affect 
the crustal equation of state \cite{GC20}.
We shall call such crust ``fully accreted'' to distinguish it from the ``partially accreted'' crust.

In what follows,
when considering the fully accreted crust we, to simplify the analysis, 
neglect small
effects of secular metric and NS radius variations
associated with the increasing mass of accreting NS \cite{GC20}.
In principle, these effects can be accounted for explicitly \cite{GC20}; 
they are, however, quite small.
In this approximation, 
the number of atomic nuclei in the fully accreted crust does not change with time. 
In other words, some mechanism of effective nuclei disintegration should operate in the inner crust.
As it is shown in \cite{GC19_DiffEq}
for smoothed compressible liquid-drop model (CLDM),
such mechanism does exist and works through the specific instability at the bottom of the inner crust,
which disintegrates nuclei into neutrons, 
so that the total number of nuclei in the crust remains constant during accretion.
Note that the actual pressure $P_{\rm inst}$ at which the instability 
sets in should not necessarily exactly coincide with $P_{\rm cc}$ 
(as CLDM of \cite{GC19_DiffEq} predicted), 
but can be smaller \cite{GC20}.
If this is the case, there is a region near the crust-core boundary,
which is decoupled from the remaining crust:
the matter is not replaced by the accreted material in that region
in the regime when the crust is fully accreted.
 
Appearance of a large number of neutrons
in the process of nuclei disintegration 
and their subsequent redistribution over the inner crust and core
results in  $P_{\rm oi}$ 
not being equal to the neutron drip pressure $P_{\rm nd}$,
at which neutrons would drip out of nuclei in the 
traditional approach 
(e.g., \cite{HZ90,Chamel_etal15_Drip}).
In particular, an excess of neutrons 
may spread across
the inner crust 
to the region, where $P<P_{\rm nd}$ \cite{GC19_DiffEq}, 
leading to the inequality $P_{\rm oi}<P_{\rm nd}$.
As a result, it turns out to be energetically favorable to capture neutrons for nuclei crossing the oi interface from the outer crust side;
this process is accompanied by substantial energy release \cite{GC20}.

We come to the following picture of the evolution
of a volume element, 
moving along with the flow of atomic nuclei 
in the
accreted crust.
Starting near the stellar surface, 
the volume element sinks under the weight of newly accreted material deeper and deeper inside the crust.
Compression induces nonequilibrium nuclear reactions in the element,
tending to bring it
to equilibrium.
In the outer crust, the number of baryons in the volume element, $N_b$, is conserved. 
Thus, 
at a given pressure,
the resulting equation of state 
can be determined by minimizing 
the Gibbs free energy {\it per baryon},
$\phi=(\varepsilon+P)/n_{b}$,
where $n_{b}$ is the baryon number density
(\cite{HZ90,HZ90b,HZ03,HZ08,Steiner12,Chamel_etal15_Drip,Fantina_ea18,SC19_MNRAS,sgc21};
see the next section for details).
Once oi interface is reached and we enter the inner crust, the situation changes 
dramatically for two reasons.
First, 
$N_b$
is not conserved anymore, 
since unbound neutrons can now freely escape from the volume element.
Second, nonequilibrium nuclear reactions induced by the compression of the element
must 
now
bring 
it
to some ``optimal'' state,
which is realized 
not only at fixed $P$ (as in the case of the outer crust),
but also [in view of Eq.\ (\ref{mun2})] at fixed $\mu_n$.
The composition
of 
the 
volume element in this state
should be determined by 
minimizing
some thermodynamic potential (let us denote it by $\Psi$)
at both $P$ and $\mu_n$ kept fixed.
The question then arises, what is this thermodynamic potential?
The next section answers this question.

\section{Thought experiment and textbook argument}
\label{exp}

The results of this section are equally applicable to both
partially and fully accreted crust.
Let us 
start with the well-known thought experiment and
put a piece of accreted crust matter 
into 
a cylindrical vessel with a massless movable piston at one of its sides. 
Assume, first, that the walls of the vessel are {\it impenetrable} for 
matter particles (nuclei, electrons, and unbound neutrons).
This assumption is equivalent to working within the traditional approach 
when modeling the NS accreted crust.

The pressure in the medium outside the vessel is $P$, the temperature is $T$.
The whole system is assumed to be always in mechanical 
and thermal equilibrium, 
so that the pressure and temperature inside the vessel are also $P$ and $T$, respectively.
Because, 
generally, the accreted crust matter is not in the full thermodynamic equilibrium,
nuclear reactions inside the vessel may proceed, 
driving the 
system
towards equilibrium.
As a result, the volume $V$ occupied by the matter in the vessel may vary
(at constant $P$ and $T$), 
and 
this process may be accompanied by the heat release and subsequent heat
transfer.
Let us calculate the transferred heat $\delta \mathcal{Q}$ 
in a short period of time,
during which 
$V$
has been changed by $d V$ 
($\delta\mathcal{Q}<0$ means that the matter in the vessel transfer heat to the outside medium). 
According to the first law of thermodynamics \cite{ll80},
\begin{align}
&
\delta \mathcal{Q} = \delta E -\delta R,
&
\label{dQ1}
\end{align}
where $\delta E$ is the change of internal energy $E$ of the matter in the vessel;
and $\delta R=-P \, d V$ is the work done on the matter by the surrounding medium.
On the other hand, in view of the second law of thermodynamics, $\delta \mathcal{Q}< T \, \delta S$, 
where $\delta S$ is the change of entropy of matter, $S$.
Combining 
these two equations, we have
\begin{align}
&
\delta E -\delta R -T \delta S<0.
&
\label{dQ2}
\end{align}
Since the process occurs at constant $P$ and $T$, 
one may introduce the thermodynamic potential (Gibbs free energy) 
$\Phi \equiv E+PV-TS$ 
and rewrite Eq.\ (\ref{dQ2}) as
\begin{align}
&
\delta\Phi<0.
&
\label{Phi}
\end{align}
This is the well-known textbook result saying that 
irreversible processes at constant $P$ and $T$ are accompanied by the decrease of $\Phi$.
Therefore, one should minimize $\Phi$ in order to determine the final state of the matter in the vessel.

Note that, under conditions typical for accreted crust, 
an approximation of vanishing temperature, $T$=0, 
is usually
well justified \cite{hpy07}. 
This approximation will be used in what follows.
In the limit $T=0$
the heat 
released in the course of nonequilibrium nuclear reactions
is simply [see Eqs.\ (\ref{dQ1})--(\ref{Phi})]:
\begin{align}
&
\delta \mathcal{Q}=\delta \Phi.
&
\label{heat}
\end{align}
In the accreted crust studies within the traditional approach one usually 
takes advantage of the fact that the total number of baryons, $N_{b}$,
in any volume element moving along with the flow of atomic nuclei
is conserved as 
the element
sinks towards the stellar center 
(in our situation, $N_{b}$ is the total number of baryons in the vessel).
This means that it is equally possible to use 
the Gibbs free energy
{\it per baryon},
$\phi \equiv \Phi/N_{b}=(\varepsilon+P)/n_{b}$, instead of $\Phi$
in all calculations
($\varepsilon \equiv E/V$).
It is the potential $\phi$, 
which is usually minimized in the traditional approach to 
the accreted crust modeling \cite{HZ90,HZ90b,HZ03,HZ08,Steiner12,Chamel_etal15_Drip,Fantina_ea18,SC19_MNRAS}.

Consider now the same thought experiment, 
but in a slightly different setup, in which 
unbound neutrons 
are present in both the vessel and outside medium and
are allowed to leak through the vessel's walls.
This assumption 
is in line with 
the nHD approach \cite{GC19_DiffEq},
in which neutrons can freely redistribute throughout the inner crust 
in order to achieve the nHD equilibrium, Eq.\ (\ref{nHDcond}).
Assume that such neutron ``leakage''  is fast enough
for the matter inside the vessel
to be
in diffusion equilibrium with respect to exchange of neutrons
with the outside medium (i.e., 
the neutron chemical potential, $\mu_{n}$,
is the same in the vessel and in the medium at all times).
What thermodynamic potential should be minimized in this situation? 

To answer this question we note, first of all,
that Eq.\ (\ref{dQ1}) for the heat release is still applicable
with the only exception that now $\delta R$ should account 
not only for the work $-P \, d V$, but also for the change of the energy of matter 
$\mu_{n} \,d N_b$ due to 
the leakage of neutrons from/to the vessel:
$\delta R= -P\, dV+\mu_{n} \, d N_{b}$.
With this redefinition of $\delta R$, Eq.\ (\ref{dQ2})
also remains correct.
Now, taking into account 
that nonequilibrium nuclear reactions in the vessel 
proceed at fixed $P$, $T$, and $\mu_{n}$, 
and introducing new thermodynamic potential, 
\begin{align}
\Psi=E +PV-\mu_{n}N_{b} -TS,
\label{Psi}
\end{align}
we immediately find that Eq.\ (\ref{dQ2}) is equivalent
to 
\begin{align}
\delta \Psi <0.
\label{Psi2}
\end{align}
We come to the very important conclusion that 
irreversible processes (nonequilibrium nuclear reactions)
inside the vessel, occurring at constant $P$, $T$, and $\mu_n$,
lead to decrease of the thermodynamic potential $\Psi$.

Again, in analogy with the case of impenetrable walls and the Gibbs potential $\Phi$,
the heat released in the vessel in the limit of $T=0$ equals
\begin{align}
\delta \mathcal{Q}=\delta \Psi.
\label{Psi3}
\end{align}
Eq. (\ref{Psi}) answers the question, posed in the end of Sec.\ \ref{nHd}
about the form of the potential that should be minimized 
in order to determine the equation of state of the inner crust.

\section{Three examples}
\label{example}

Below we consider three examples in which $\Psi$ can be calculated.
We assume that $T=0$ in all these examples.
The first example is the ground-state crust. 
In this case, Eq.\ (\ref{Psi}) gives $\Psi=0$, 
because in the ground state one has $\varepsilon+P=\mu_{n} n_{b}$ \cite{hpy07}. 

The second example is the smoothed CLDM described in \cite{GC19_DiffEq}.
Using it, \cite{GC19_DiffEq} constructed 
the 
first 
thermodynamically
consistent equation of state for an accreted crust, 
satisfying the condition (\ref{nHDcond}).
In the smoothed CLDM
the atomic mass number $A$ and charge number $Z$ are treated as continuous variables.
Thus, 
at each pressure $P$, 
nuclei of only one species $(A,Z)$ are present (no mixtures).
The energy density in CLDM depends on the two parameters, $n_{b}$ and the number density
of atomic nuclei, $n_{\rm N}$.
The first law of thermodynamics for this model reads
\begin{align}
d \varepsilon = \mu_{n} d n_{b} + \mu_{\rm N} d n_{\rm N},
\label{2nd}
\end{align}
where $\mu_{\rm N}$ is the effective chemical potential describing the energy change due to 
creation of additional nuclear cluster in the system at fixed $n_{b}$.
In turn, the pressure $P$ is given by [c.f. Eq.\ (\ref{Pres}) below]: 
$P=-\varepsilon+\mu_{n}n_{b}+\mu_{\rm N}n_{\rm N}$.
Using the latter formula, one finds 
\begin{align}
\Psi=\mu_{\rm N} N_{\rm N},
\label{Psi4}
\end{align}
where $N_{\rm N}$ is the total number of nuclei in a given volume element moving along with the nuclei.
As follows from the results of \cite{GC19_DiffEq},
$N_{\rm N}$ remains constant as the volume element sinks towards the crust-core boundary.
Moreover, it can be shown \cite{GC19_DiffEq} that Eqs.\ (\ref{nHDcond}) and (\ref{hydro}) imply that
the redshifted $\mu_{\rm N}$ should be constant
throughout the inner crust, $\mu_{\rm N}^\infty={\rm const}$.
Combining these two properties, 
one finds
\begin{align}
\Psi^\infty={\rm const} \quad \quad {\rm or} \quad \quad \psi^\infty={\rm const},
\label{Psi5}
\end{align}
where in the right formula we introduced the potential $\Psi$ {\it per nucleus}: $\psi \equiv \Psi/N_{\rm N}$.
Eq.\ (\ref{Psi5}) does not, in fact,  
rely on
a particular nuclear model (CLDM, in our case)
and is quite general, 
as is shown below in this section.

In the third example,
let us 
assume that the crust matter 
is composed of a mixture of nuclei of different species $j$
and, possibly, unbound neutrons 
(for simplicity, we 
ignore unbound protons, 
which could be present near the crust-core 
transition
\cite{pcpfddg18}).
The first law of thermodynamics for the mixture can then be written,
in analogy to equation (\ref{2nd}), as
\begin{align}
&
d \varepsilon =\mu_{n} d n_{b}+\sum_j \mu_{\rm N}^{(j)} d n_{\rm N}^{(j)},
&
\label{1st4}
\end{align}
where $n_{\rm N}^{(j)}$ and $\mu_{\rm N}^{(j)}$
are, respectively, the number density and effective chemical potential for nucleus species $j$ 
($\mu_{\rm N}^{(j)}$ is analogous to $\mu_{\rm N}$ from the previous example).
Introducing the total number density $n_{\rm N}$ of nuclei,
the fraction $x_{\rm N}^{(j)}$ of nucleus species $j$ ($\sum_j x_{\rm N}^{(j)}=1$),
and the average chemical potential $\mu_{\rm N}$,
\begin{align}
&
n_{\rm N} \equiv \sum_j n_{\rm N}^{(j)}, \quad  x_{\rm N}^{(j)} \equiv\frac{n_{\rm N}^{(j)}}{n_{\rm N}},
\quad 
\mu_{\rm N} \equiv \sum_j \mu_{\rm N}^{(j)} x_{\rm N}^{(j)},
&
\label{nNxN}
\end{align}
Eq.\ (\ref{1st4}) can be represented in the form
\begin{align}
&
d \varepsilon =\mu_{n} d n_{b}
+\mu_{\rm N} d n_{\rm N}+\sum_j \mu_{{\rm N}}^{(j)} n_{\rm N} d x_{{\rm N}}^{(j)}.
&
\label{1st5}
\end{align}
The pressure is defined in a standard way,
\begin{align}
&
P=- \frac{\partial (\varepsilon V)}{\partial V}
=-\varepsilon + \mu_{n} n_{b} +\mu_{\rm N} n_{\rm N},
&
\label{Pres}
\end{align}
where the partial derivative is taken at fixed $N_{\rm N}^{(j)}=n_{\rm N}^{(j)}V$ 
and $N_{b}=n_{b} V$
($j$ runs over all nucleus species).
It is clear from (\ref{Psi}) and (\ref{Pres}) that the thermodynamic potential $\Psi$ 
for the mixture is given by the same expression (\ref{Psi4}) as in the CLDM,
but with $\mu_{\rm N}$ defined by Eq.\ (\ref{nNxN}) and
$N_{\rm N}=\sum_{j} N_{\rm N}^{(j)}$.

Now, let us assume that 
the crust is fully accreted and
there are no nuclear reactions 
in some region of the inner crust.
This means that 
$N_{\rm N}={\rm const}$
and $x_{\rm N}^{(j)}={\rm const}$ 
in an arbitrary chosen volume element, moving downward with the nuclei in that region.
Consequently, equation (\ref{1st5}) reduces to (\ref{2nd}).
Then, using the nHD condition (\ref{nHDcond}), as well as Eqs.\ (\ref{hydro}), (\ref{2nd}), and (\ref{Pres})
it is straightforward to demonstrate (see also \cite{GC19_DiffEq}) 
that there must be $\mu_{\rm N}^\infty={\rm const}$, 
and hence (because $N_{\rm N}={\rm const}$), 
$\Psi^\infty={\rm const}$.
We come to the conclusion that Eq.\ (\ref{Psi5}) 
is satisfied in the region of the inner crust, in which there are no nuclear reactions.
Very similar arguments can be applied to the outer crust and thermodynamic potential $\Phi$.
Namely, it can be shown (see also Supplemental material in \cite{GC20}) that 
\begin{align}
\Phi^\infty={\rm const} \quad \quad {\rm or} \quad \quad \phi^\infty={\rm const}
\label{Phi2}
\end{align}
in those regions of the outer crust, 
in which nuclear composition is fixed.
To derive (\ref{Phi2}) it should be noted 
that in the outer crust $n_{b}=\langle A\rangle n_{\rm N}$, 
where $\langle A\rangle$ is the average atomic mass number in the mixture.

The consideration above can also be applied to a region of partially accreted crust,
for which $x_{\rm N}^{(j)}={\rm const}$ 
for all types of nuclei.
For example, if the region is situated in the inner crust, 
then it follows from Eqs.\ (\ref{nHDcond}),
(\ref{hydro}), (\ref{2nd}), and (\ref{Pres})
that it must be $\mu_{\rm N}^\infty=\psi^\infty={\rm const}$ in that region.
In the outer crust the latter condition should be replaced with $\phi^\infty={\rm const}$.

\section{Heat release in the crust}
\label{heat2}
The results obtained above can be utilized for calculation 
of the heat 
release (and its distribution)
in the 
fully accreted
crust in the course of accretion.
Let us consider two points, I and II,
with the pressure $P_{\rm I}$ and $P_{\rm II}$,
located close to each other 
in the 
inner crust of an NS.
Consider now a small volume element in point I, which moves with the nuclei.
The thermodynamic potential of this element is $\Psi=\Psi_{\rm I}$.
After a while, the element will shift to point II, where $\Psi$ will be equal to $\Psi_{\rm II}$. 
What is the value of the heat released in such process?

The heat will be generated 
by 
nonequilibrium nuclear reactions,
which can occur in the volume element 
due to the variation of pressure from $P_{\rm I}$ to $P_{\rm II}$.
To calculate the heat, let us assume that the process takes place in two steps:
(i) first, the volume element moves from point I to point II with frozen composition (no heat is generated!);
(ii) then, in point II, all the required nuclear reactions occur and nuclear composition modifies.%
%
\footnote{In this picture, the smooth variation of matter composition in the crust
is approximated as an infinite set of infinitely small phase transitions 
at which the composition changes in a step-wise manner.}
%
In view of Eq.\ (\ref{Psi5}), we can write for the step (i):
\begin{align}
&
\Psi_{\rm I} \, {\rm e}^{\nu_{\rm I}/2} =
\Psi_{\rm II, 0}\, {\rm e}^{\nu_{\rm II}/2},
&
\label{PsiIII}
\end{align}
where ${\rm e}^{\nu_{\rm I}/2}$ and ${\rm e}^{\nu_{\rm II}/2}$
are the redshifts at points I and II, respectively;
$\Psi_{\rm II,0}$ 
is the thermodynamic potential of the volume element 
at point II 
{\it before} nuclear reactions were initiated.
At the step (ii) the nonequilibrium nuclear reactions 
change $\Psi$ from $\Psi_{\rm II,0}$ to $\Psi_{\rm II}$
at constant  $P$ and $\mu_{n}$.
Because of (\ref{Psi3}), the heat $\delta Q$ {\it released} in the system,
equals:%
%
\footnote{Part of this energy can be carried away from the star by neutrinos, 
see, e.g., \cite{Gupta_ea07}.}
%
$\delta Q =\Psi_{\rm II,0}-\Psi_{\rm II}$
[the sign differs from that adopted in Eq.\ (\ref{Psi3}), 
because in (\ref{Psi3}) the emitted heat is, by definition, negative,
which is not convenient for us in what follows].
Accounting for (\ref{PsiIII}), 
the redshifted heat release is presented as
\begin{align}
&
\delta Q^\infty=(\Psi_{\rm II,0}-\Psi_{\rm II}) \, {\rm e}^{\nu_{\rm II}/2}
=\Psi_{\rm I}^\infty-\Psi_{\rm II}^\infty.
&
\label{Q}
\end{align}
If points I and II are infinitely close to each other, the heat released in the inner crust equals:
\begin{align}
&
\delta Q^\infty=-\frac{d\Psi^\infty}{d P} \, dP,
&
\label{Q2}
\end{align}
where we parametrize $\Psi^\infty$ in the fully accreted crust as a function of $P$.
A similar formula can be derived for the heat release in the outer crust,
\begin{align}
&
\delta Q^\infty=-\frac{d\Phi^\infty}{d P} \, d P,
&
\label{Q3}
\end{align}
with the potential $\Phi$ instead of $\Psi$.%
%
\footnote{In the case of strong phase transitions with a finite amount of heat release, 
the derivatives in 
Eqs.\ (\ref{Q2}) and (\ref{Q3}) equal the corresponding delta-functions with the weights 
$\Psi_{\rm I}^\infty-\Psi_{\rm II}^\infty$ and $\Phi_{\rm I}^\infty-\Phi_{\rm II}^\infty$, respectively.
}
%
Eqs.\ (\ref{Q2}) and (\ref{Q3}) give the {\it distribution} of the redshifted 
heat 
generated in the fully accreted crust of an NS.
The total redshifted 
heat 
release, $Q^\infty$, 
in the fully accreted crust can be obtained by integrating 
$\delta Q^\infty$ over the whole crustal volume, 
where the heat is generated, i.e., 
from $P=P_{\rm ash}$ to $P=P_{\rm inst}$
($P_{\rm ash}$ is the pressure near the stellar surface, 
at which the accreted material has already 
fused into heavy nuclei).
When doing this, one should be careful with 
the evaluation of 
the redshifted heat release 
$\Delta Q_{\rm oi}^\infty$ and $\Delta Q_{\rm inst}^\infty$ at,
respectively, 
the oi interface 
and
the bottom of the crust ($P=P_{\rm inst}$), 
where nuclei disintegrate into neutrons.
All in all, one can write:
\begin{align}
&
Q^\infty =- \int_{P_{\rm ash}}^{P_{\rm oi}}
\frac{d\Phi^\infty}{d P} \, d P 
+ \Delta Q_{\rm oi}^\infty
-\int_{P_{\rm oi}}^{P_{\rm inst}}
\frac{d\Psi^\infty}{d P} \, d P + \Delta Q_{\rm inst}^\infty
&
\nonumber\\
&
= \Phi^\infty(P_{\rm ash})-\Phi^\infty(P_{\rm oi})+\Delta Q_{\rm oi}^\infty 
\nonumber\\
&
+ \Psi^\infty(P_{\rm oi})-\Psi^\infty(P_{\rm inst})+ \Delta Q_{\rm inst}^\infty.
&
\label{Q4}
\end{align}
Let us first calculate the heat release $\Delta Q_{\rm inst}^\infty$
due to complete disintegration of atomic nuclei at the bottom of the crust.
Because disintegration is complete (all nuclei from the given volume element disintegrate,
hence $\Psi^\infty=0$ in the final state)
and occurs at fixed $P=P_{\rm inst}$ and $\mu_n$,
the heat release is given, in accordance with Eqs.\ (\ref{Psi3}) and (\ref{Psi4}),
by the formula:
$\Delta Q_{\rm inst}^\infty=\Psi^\infty(P_{\rm inst})$.

To calculate $\Delta Q_{\rm oi}^\infty$, 
let us consider a volume element attached to nuclei 
and initially situated near the oi interface on the outer crust side.
Its redshifted 
Gibbs free energy
equals $\Phi^\infty=\Phi^\infty(P_{\rm oi})$.
After crossing the oi interface, the element enters the inner crust,
where unbound neutrons with the chemical potential  $\mu_{n, {\rm oi}}=m_{n}$ are present.
This will initiate the neutron capture reactions, 
mentioned in Sec.\ \ref{nHd} (see also \cite{GC20,sgc21}),
which will lead to the heat release $\Delta Q_{\rm oi}^\infty$.
According to Eq.\ (\ref{Psi3}), it equals
$\Delta Q_{\rm oi}^\infty=\Psi^\infty_{0}(P_{\rm oi})-\Psi^\infty(P_{\rm oi})$,
where $\Psi^\infty_{0}(P_{\rm oi})$
is the redshifted thermodynamic potential $\Psi^\infty$ 
of the volume element, immediately after it crosses the oi interface;
and $\Psi^\infty(P_{\rm oi})$ is the 
corresponding potential after 
all reactions in the volume element proceed
[it coincides with $\Psi^\infty(P_{\rm oi})$ in Eq.\ (\ref{Q4})].
In view of the definition of $\Phi$ and Eq.\ (\ref{Psi}), 
one can write:
$\Psi^\infty_{0}(P_{\rm oi})
=\Phi^\infty(P_{\rm oi})-\mu^\infty_{n,{\rm oi}} N_{b}$,
where $N_{b}$ is the number of baryons in this element {\it before} it crossed the oi interface.
Collecting the above equations, we get
\begin{align}
&
\Delta Q_{\rm oi}^\infty
=\Phi^\infty(P_{\rm oi})-\mu^\infty_{n,{\rm oi}} N_{b}-\Psi^\infty(P_{\rm oi}).
&
\label{Q6}
\end{align}
Note that, due to the condition (\ref{nHDcond}),
$\mu^\infty_{n,{\rm oi}}$
equals $\mu_{n}^\infty$ at any point of the inner crust and core.
On the other hand, 
in the core $\mu_n$ coincides with the baryon chemical potential, 
$\mu_{b,{\rm core}}$ \cite{hpy07} (we assume that the matter in the core is in beta-equilibrium).
Thus, $\mu^\infty_{n,{\rm oi}}=\mu_{b,{\rm core}}^\infty$.
Using this formula together with the expressions 
$\Delta Q_{\rm inst}^\infty=\Psi^\infty(P_{\rm inst})$
and (\ref{Q6}), 
one can rewrite (\ref{Q4}) as (see also \cite{GC20})
\begin{align}
&
Q^\infty=\Phi^\infty(P_{\rm ash})-\mu_{b,{\rm core}}^\infty N_{b}.
&
\label{Q7}
\end{align}
In the fully accreted crust, neglecting small secular metric and radius
variations, the number of accreted baryons per unit time equals the
number of baryons crossing the oi interface per unit time \cite{GC20}.
Thus, dividing (\ref{Q7}) by $N_{b}$, and using the fact that 
$\Phi^\infty(P_{\rm ash})\approx \overline{m}_{b, {\rm ash}} N_{b} {\rm e}^{\nu_{\rm s}/2}$, 
where $\overline{m}_{b, {\rm ash}}$ is the average mass of ashes per baryon and 
${\rm e}^{\nu_{\rm s}/2}$ is the redshift factor at the stellar surface,
we arrive at the final formula for the redshifted heat release $q^\infty$ per accreted baryon:
\begin{align}
&
q^\infty= \overline{m}_{b, {\rm ash}} {\rm e}^{\nu_{\rm s}/2}-\mu_{b,{\rm core}}^\infty.
&
\label{Q8}
\end{align}
This formula coincides with the similar formula (3) from \cite{GC20}
derived using a different method.

\section{Conclusion}
\label{concl}

We find the thermodynamic potential $\Psi$ [Eq.\ (\ref{Psi})],
that should be minimized in order to derive 
the
equation of state 
for
the inner crust of accreting NS.
This potential differs from the Gibbs thermodynamic potential $\Phi$,
adopted in the previous studies 
(e.g., \cite{HZ90,HZ90b,HZ03,HZ08,Steiner12,Chamel_etal15_Drip,Fantina_ea18,SC19_MNRAS}), 
which were carried out in the traditional approach,
ignoring the condition (\ref{nHDcond}).
If the total number of atomic nuclei $N_{\rm N}$
in a volume element moving along with the nuclei
in some region of the inner crust is conserved,
it is sufficient to minimize the potential $\Psi$ per one nucleus,
$\psi=\Psi/N_{\rm N}=(\varepsilon+P-\mu_{\rm n} n_{b})/n_{\rm N}$, 
where $n_{\rm N}$ is the total number density of 
nuclei.

We also show how to find the {\it heat release distribution} 
due to nonequilibrium reactions
in the fully accreted crust
as a function of $P$, 
once the potential $\Psi$ in the inner crust (or $\Phi$ in the outer crust) is known
[see Sec.\ \ref{heat2}, 
particularly, Eqs.\ (\ref{Q2}) and (\ref{Q3})].
This allows us to derive the formula (\ref{Q8}) 
for the total heat release
per baryon, 
which
coincides with the similar formula 
derived 
in a completely different way in \cite{GC20} (their formula 3).

The results obtained in this Letter
provide a theoretical basis
for the adequate description 
of 
the 
accreted NS crust.
We believe 
that the approach developed here
may prove to be especially useful in applications to the 
partially accreted crust,
when the crustal equation of state changes with time.
Finally, we emphasize that the potential $\Psi$ introduced in this paper
can (and should) be used in other problems,
dealing with particle mixtures at fixed pressure and chemical potential of one of the particle species. 
Generalization of $\Psi$ to the case when a few chemical potentials 
are kept fixed 
simultaneously,
is straightforward.

\begin{acknowledgments}
MEG is partially supported 
by RFBR [Grant No.\ 19-52-12013]. 
\end{acknowledgments}


%

\end{document}